\documentclass{aa}

\usepackage{graphicx}
\usepackage{subfig}
\usepackage{txfonts}
\newcommand{\oii}{[O\,{\sc{ii}}]}
\newcommand{\oiii}{[O\,{\sc{iii}}]}
\def\ha{H$\alpha$}

\begin{document}

   \title{The OTELO Survey: The main sequence of low-mass galaxies}

   \author{Bernab\'e Cedr\'es
          \inst{1,2,3,4}
          \and
          Jordi Cepa
          \inst{1,2,4}
          \and
          Carmen P. Padilla--Torres
          \inst{1,2,5,4}
          \and
          \'Angel Bongiovanni
          \inst{6,4}
          \and
          Miguel Cervi\~no
          \inst{7}
          \and
          Jos\'e A. de Diego
          \inst{8}
          \and
          Jakub Nadolny
          \inst{9,1,2}
          \and
          Simon B. De Daniloff
          \inst{6,10}
          \and
          Mauro Gonz\'alez--Otero
          \inst{11,4}
          \and
          Monica I. Rodr\'iguez
          \inst{6}
          \and
          Jes\'us Gallego
          \inst{12}
          \and
          Mirjana Povi\'c
          \inst{11,13,14}
          \and
          Maritza A. Lara--L\'opez
          \inst{12}
          \and
          J. Ignacio Gonz\'alez--Serrano
          \inst{15}
          \and
          Miguel S\'anchez--Portal
          \inst{6}
          \and
          Ana Mar\'ia P\'erez--Garc\'ia
          \inst{16,4}
          \and
          Ricardo P\'erez--Mart\'inez
          \inst{16,4}
          \and
          Emilio J. Alfaro
          \inst{11}
          }

   \institute{Instituto de Astrof{\'i}sica de Canarias (IAC), 38200 La Laguna, Tenerife, Spain\\
              \email{bcedrese@professor.universidadviu.com}
         \and
             Departamento de Astrof{\'i}sica, Universidad de La Laguna (ULL), 38205 La Laguna, Tenerife, Spain
        \and
            Universidad Internacional de Valencia (VIU), C/Pintor Sorolla 21, E-46002 Valencia, Spain
        \and
            Asociaci\'on Astrof{\'i}sica para la Promoci\'on de la Investigaci\'on, Instrumentaci\'on y su Desarrollo, ASPID, 38205 La Laguna, Tenerife, Spain
        \and
            Fundaci\'on Galileo Galilei--INAF, Rambla Jos\'e Ana Fern\'andez P\'erez, 7, 38712 Bre\~na Baja, Tenerife, Spain
        \and
            Institut de Radioastronomie Millim\'etrique (IRAM), Av. Divina Pastora 7, N\'ucleo Central 18012, Granada, Spain  
        \and
            Centro de Astrobiolog\'ia (CAB), CSIC-INTA, Camino Bajo del Castillo s/n, 28692, Villanueva de la Ca\~nada, Madrid, Spain
        \and
            Instituto de Astronom\'ia, Universidad Nacional Aut\'onoma de M\'exico, A.P. 70-264, 04510 M\'exico D. F., M\'exico
        \and
            Astronomical Observatory Institute, Faculty of Physics and Astronomy, Adam Mickiewicz University, ul.~S{\l}oneczna 36, 60-286 Pozna{\'n}, Poland
        \and
            Dpto de F\'isica Te\'orica y del Cosmos, Edificio Mecenas, Campus de Fuentenueva, Universidad de Granada, E-18071, Granada, Spain
        \and
            Instituto de Astrof\'isica de Andaluc\'ia–-CSIC, Glorieta de la Astronom\'ia s/n, E–18008 Granada, Spain
        \and
            Departamento de F\'isica de la Tierra y Astrof\'isica, Instituto de F\'isica de Part\'iculas y del Cosmos, IPARCOS, Universidad Complutense de Madrid (UCM), 28040 Madrid, Spain
        \and
            Space Science and Geospatial Institute (SSGI), Entoto Observatory and Research Center (EORC), Astronomy and Astrophysics Research Division, PO Box 33679, Addis Abbaba, Ethiopia
        \and
            Physics Department, Mbarara University of Science and Technology (MUST), Mbarara, Uganda
        \and
            Instituto de F\'isica de Cantabria (CSIC-Universidad de Cantabria), 39005 Santander, Spain
        \and
            ISDEFE for European Space Astronomy Centre (ESAC)/ESA, PO Box 78, 28690 Villanueva de la Ca\~nada, Madrid, Spain
             }

   \date{}

  \abstract
   {}
   {This paper describes the relationship between the star formation rate and the stellar mass, the so-called main sequence (MS), of low-mass galaxies at three different redshifts from the OTELO survey. In particular, we study \ha\ at $z=0.38$, \oiii\ at $z=0.83$, and \oii\ at $z=1.43$ emission line galaxies (ELGs). 
   This is done to characterise the properties of low-mass ELGs.}
   {We fitted the spectral energy distribution (SED) of each emitter and obtained the stellar masses, the total infrared luminosity, and the star formation rate. 
   We found that 100\% of the \ha\ emitters and about 90\% of the \oiii\ and \oii\ emitters are low-mass galaxies ($<10^{10}\mathrm{M}_{\odot}$). We generated a MS for each redshift regime, employing all galaxies and binning them in stellar mass. We obtained the parameters of the fit (turnover mass and slope of the low-mass regime) and compared them to results from the literature.}
   {We found that the colour--magnitude method employed to select ELGs leaves out a significant number of them. We also found that the whole sample contains few luminous infrared galaxies and no ultra-luminous infrared galaxies. We also found that the lack of infrared constraints in the input SEDs may generate problems when determining the total infrared luminosity. When comparing the MS obtained for our sample of low-mass ELGs with those from the bibliography, we found, in general,  very good agreement. We can consider the MS to be almost the same for the relatively unexplored regime of low-mass galaxies. However, the turnover mass obtained by this work is higher for all redshifts compared with the ones from the literature. We suggest that this is an effect of the different methods and samples used in the generation of the MS compared to those of the bibliography.}
   {}


   \maketitle
%
%-------------------------------------------------------------------

\section{Introduction}

The galaxy main sequence (MS) of star formation is a relationship between the stellar mass ($M_*$) of galaxies and their star formation rate (SFR). It was presented for local galaxies by \cite{brinchmann2004}. Since then, it has been extended to higher redshifts by several authors (see, for example, \citealt{noeske2007} or \citealt{daddi2007}; and for more recent works \citealt{popesso2023}, \citealt{euclid2025}, and references therein).

The MS relationship is pretty tight, with a 0.2\,dex width. Only a select number of star-forming galaxies (SFGs) lie outside the main body. \cite{elbaz2011} found that the majority of the MS was composed of ‘normal’ SFGs, and that a population of high starburst galaxies with high specific star formation rates (sSFRs) fell above the MS. \cite{rodighiero2011} showed that only a fraction of SFGs undergo merger events that should increase the SFR of the galaxies, so these galaxies are the ones that are part of the outliers in the MS (\citealt{lee2015}). Taking this into account, the MS relationship seems to indicate that galaxies undergo a regular star formation history, in opposition to a history driven by mergers and starbursts (\citealt{popesso2019}).

\cite{noeske2007} and \cite{speagle2014} proposed a functional form for the MS as a single power law. However, later on, with more data available, \cite{lee2015} proposed a more complex form, given by
\begin{equation}
    SFR(M_*)=SFR_{max}/\left(1+(M_*/M_0)^{-\gamma}\right),
    \label{eclee}
\end{equation}
where $SFR_\mathrm{max}$ is the saturation value for the SFR, $M_0$ is the turnover mass, and $\gamma$ is the slope for the low-mass regime. \cite{lee2015} gave a range of $0.9\leq\gamma\leq1.3$ for the slope of the low-mass regime, and a turnover mass of about $M_0\simeq10^{10}$M$_{\odot}$, with some evolution at higher redshifts. \cite{popesso2023} compiled a large set of data on SFGs from the literature in the redshift range $0<z<6$, and found strong evidence of a linear evolution of the turnover mass with time, as well as the saturation value. They found no evidence of $\gamma$ evolution and obtained a value compatible with 1. The turnover mass is discussed there as determining the shape of the MS. For stellar masses well below the turnover mass, the galaxies have a nearly constant sSFR. In addition, for stellar masses larger than the turnover mass, the SFR is suppressed. Several authors (\citealt{popesso2019}, \citealt{tacchela2016}, \citealt{daddi2022}, and references cited therein) found that the MS was dominated by central galaxies (the main galaxy occupying the dark matter haloes). \cite{popesso2023} extend this relation to all redshifts and then interpret the turnover mass as defining the transition between an environment that sustains the star-forming processes for the central galaxy to a regime in which the star formation is suppressed.

On the other hand, the work on low-mass galaxies ($<10^{10}\mathrm{M}_{\odot}$) is less extensive. For example, \cite{terao2022} found in the MS of a sample of \ha\ emitters from $2.1<z<2.5$ that low-mass galaxies below their completeness limit present high sSFR values. They explained this, suggesting that the low-mass galaxies experienced a recent shorter-timescale starburst compared with larger galaxies. Along the same lines, \cite{chen2024} in $z\sim2.3$ found that \ha-derived SFRs may be more sensitive to short recent starbursts than other indicators, and this may explain the scatter on the MS obtained by the simulations by \cite{sparre2017}. In opposition, the work of \cite{schreiber2015} found that low-mass galaxies presented the same dispersion as galaxies with moderately high masses. More recently, several works have implied that this dispersion may be due to the incompleteness of the data for low-mass galaxies. For example, \cite{simmonds2025}, employing data from JADES (JWST Advanced Deep Extragalactic Survey), found that the sSFR is independent of stellar mass when the completeness of the sample is taken into account.

Here, we study the evolution of the MS with the redshift for a sample of low-mass emission line galaxies (ELGs) from the OTELO (OSIRIS Tunable Filter Emission Line Object) Survey. Since the low-mass regime for galaxies of MS is not well studied, comparisons with galaxies of intermediate- to high-mass regimes could help us to obtain a clearer picture of the true form of the MS. The paper is organised as follows. In Sect. \ref{sec2} we describe the OTELO Survey and the three subsets of ELGs, selected covering the range $0.38\leq z \leq 1.43$ in redshift. In Sect. \ref{sec3} we fit the spectral energy distribution (SED) of the ELGs through the CIGALE code and study several derived parameters such as the stellar mass, the total infrared luminosity, and the SFR. In Sect. \ref{sec4} we fit the derived stellar masses and the SFRs to eq. \ref{eclee}, to obtain the MS, and we show the comparison with the data available in the literature. In Sect. \ref{sec5} we present the main conclusions of this work.

Throughout this paper, we assume a standard $\Lambda$ cold dark matter cosmology with $\Omega_\Lambda = 0.7$, $\Omega_{\mathrm{m}} = 0.3$, and $\mathrm{H}_0 = 70 \,\mathrm{km}\,  \mathrm{s}^{-1}\, \mathrm{Mpc}^{-1}$. We also assume a \cite{chabrier2003} initial mass function (IMF), unless otherwise stated.

\section{The OTELO Survey} \label{sec2}
The OTELO Survey is a blind, pencil-beam survey that covers a region of 56\,arcmin$^2$ of the Extended Groth Field. It was observed with the OSIRIS instrument, located at Gran Telescopio Canarias (GTC). The survey aims to detect galaxies with emission lines. It uses 2D spectroscopy techniques, employing the tunable filter mode of the instrument, with a tomography of 36 slices evenly distributed, covering an atmospheric window between 8950 and 9300 $\AA$, with a spectral resolution of $R\sim700$. The catalogue contains 11237 sources and is 50\% complete with an AB magnitude of 26.38. The catalogue also includes a pseudo-spectrum for each object detected. This pseudo-spectrum is the result of the convolution of the object SED by the response of the OSIRIS instrument. The catalogue also contains ancillary photometric data that cover from the X-ray to far-infrared range. The survey, its data products, and reduction processes are described in detail in \cite{otelo}\footnote{The catalogue can be accessed at http://research.iac.es/proyecto/otelo/pages/otelo.php}.
The photometric redshifts for all objects were obtained by fitting the data to SED templates using the LePhare code (\citealt{arnouts1999}, \citealt{ilbert2006}).

From the OTELO Survey, we selected three subsets of sources based on their emission in \ha\ at $\langle z \rangle =0.38$, with 34 emitters (\citealt{marina2019}), \oiii\ at $\langle z \rangle=0.83$, with 184 emitters (\citealt{bongio2020}), and \oii\ at $\langle z \rangle=1.43$, with 60 emitters (\citealt{cedres2021}). The selection criteria used for each subset were based, first, on locating all objects with the appropriate photometric redshift, and then on visually inspecting each pseudo-spectrum for emission by several researchers. The specific details of the selection process for each subset are explained in the reference articles (\citealt{marina2019}, \citealt{bongio2020}, and \citealt{cedres2021} for \ha, \oiii, and \oii\ , respectively). The spectroscopic redshift, as well as the integrated fluxes for the lines and the continua, was obtained for all galaxies in the subsamples, calculated through the inverse convolution method, developed in \cite{kuba2020}. In short, a series of Monte Carlo simulations were performed ($\sim10^6$), simulating the redshift, the flux of the continuum, the dispersion of the line, and the flux of the line; then, each simulation was convolved with the response of the tunable filter system and a set of synthetic pseudospectra were obtained. After a  $\chi^2$ minimisation procedure, the parameters mentioned above were obtained.

Regarding possible contamination of the subsets by the AGNs, \cite{rocio2021} found a very low fraction ($\sim2\%$) of AGNs in their emitter sample at $z\sim0.9$. They explained this by the fact that OTELO is a very deep survey, but covers a small volume, and taking into account that the AGN fraction depends more on luminosity than redshift (\citealt{chiang2019}), the general low luminosity of the emitters (linked to the low masses) implies a negligible number of AGNs. On the same note, \cite{marina2019} found no galaxies hosting AGNs at low luminosities, and \cite{cedres2021} concluded, based on assumptions by \cite{drake2013}, that the possible AGN contamination of their sample at $z\sim1.43$ was about $6\%$, and together with the selection criteria that the subsets comprise only galaxies fitted by a SED corresponding to templates belonging to a starburst or a star-forming galaxy. Taking all this into account, we may assume that all subsets have little contamination by AGNs.

\begin{figure}
    \centering
    \includegraphics[width=\hsize]{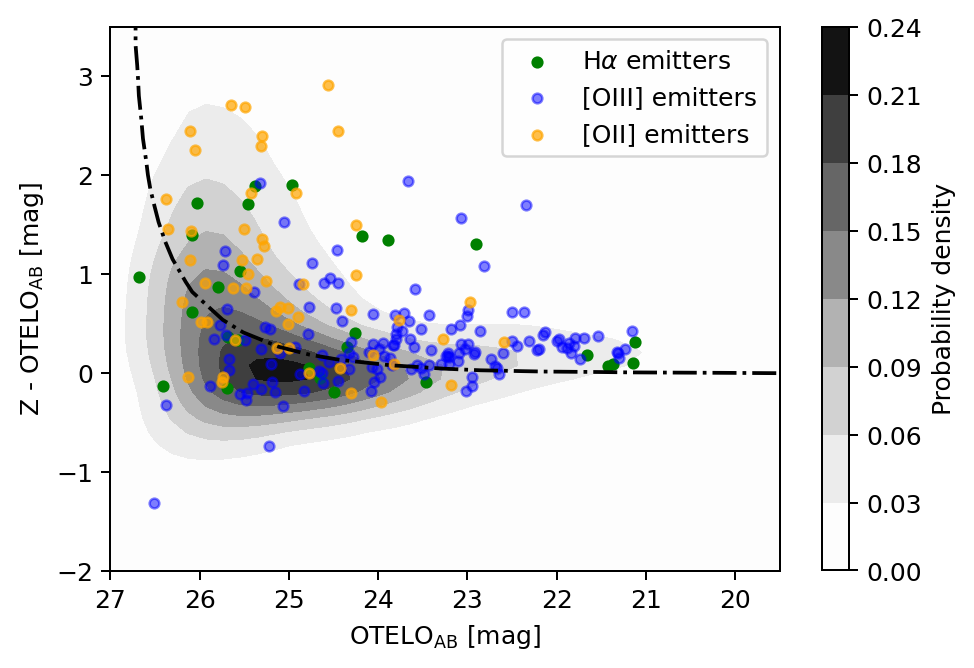}
    \caption{Z-OTELO$_{AB}$ colour--magnitude diagram. The filled grey contours represent the density of all the OTELO sources. The \ha, \oiii, and \oii\ emitters are presented by green, blue, and orange dots, respectively. The dash-dotted black line is the $2\Sigma$ isoline of colour significance from \cite{otelo}.}
    \label{fig:color}
\end{figure}

In Fig. \ref{fig:color} we show a colour--magnitude diagram from all OTELO sources, represented as density following the grey scale of the colour bar,  and the selected emitters are drawn as yellow, red, and black dots for \ha, \oiii\ , and \oii\ emitters, respectively. There, the OTELO$_{\mathrm{AB}}$ magnitudes are the instrumental fluxes measured in the OTELO--Deep image (resulting from the weighted combination of all the slices of the tomography obtained by the tunable filter) and converted into AB magnitudes (see \citealt{otelo}).
The dash-dotted line represents the $2\Sigma$ isoline of colour-excess significance obtained after \cite{bunker1995} and also presented in \cite{otelo}. There, it is defined as
\begin{equation}
   (Z-\mathrm{OTELO_{AB}})=-2.5\log\left[ 1-\delta\Sigma10^{-0.4(zp-\mathrm{OTELO_{AB}})}\right],
\end{equation}
where $zp=30.504$ is the photometric zero-point of $\mathrm{OTELO_{AB}}$ data, $\delta$ is the sum in quadrature of the sky background in each band, and $\Sigma$ is the colour excess significance.
This line was used as a discriminant for the ELGs in the OTELO Survey, where all sources with $\Sigma>2$ and a signal-to-noise ratio on the continuum better than $5\sigma$ were selected as emitters. It can be seen that most of the emitters from the subsets are located above the $2\Sigma$ line, but there is a non-negligible number of emitters from all subsets below the line. This happens because the emitters employed in this study were selected by visual inspection of the pseudo-spectrum of pre-selected candidates (a detailed report of the method is described in \citealt{cedres2021}). This shows that employing only colour--magnitude relationship techniques in the selection of ELGs will rule out an important number of them; in this case, it is $\sim$29\% of all the emitters in our sample. This limitation in the detection of emitters was also reported in \cite{otelo}.

\section{CIGALE fits} \label{sec3}

In order to study the physical properties of the selected subsamples, we performed a SED fitting using CIGALE (\citealt{cigale}). The bands used in the fittings were the ones included in the ancillary data from the core catalogue, as well as the ones from the multi-wavelength catalogue, when available, from the OTELO survey (\citealt{otelo}), i.e.: u, g, i, and z from the CFTHLS survey; F606W and F814W from HST-ACS; the J, H, and K$_{s}$ bands from WIRDS; the far-UV and near-UV from Galaxy Evolution Explorer (GALEX, \citealt{martin2005}); 3.6, 4.5, 5.8, and 8.0 $\mu$m from the IRAC instrument and 24 $\mu$m from the MIPS instrument at Spitzer Space Observatory; 100 and 160 $\mu$m from the PACS instrument; and 250, 300, and 500 $\mu$m from the SPIRE instrument at Herschel Space Observatory. However, not all bands were available for the full subsamples. In Table \ref{tab:infra} we summarise the total number of objects with emission detected in the IR bands.

\begin{table*}[]
    
    \centering
    \caption{Percentage of emitters with data in each IR band.}
    \begin{tabular}{c|c|c|c|c|c|c|c|c|c|c}
    \hline\hline
    Subsample  &  IRAC  & IRAC  & IRAC & IRAC  & MIPS  & PACS  & PACS  & SPIRE  & SPIRE  & SPIRE \\
     & 3.5$\mu$m & 4.5$\mu$m & 5.8$\mu$m & 8.0$\mu$m & 24$\mu$m & 100$\mu$m & 160$\mu$m & 250$\mu$m & 300$\mu$m & 500$\mu$m\\
    \hline
    \ha & 20.5\% & 17.6\% & 14.7\% & 14.7\% & 14.7\% & 0\% & 0\% & 3.0\% & 3.0\% & 0\%\\
    \oiii & 52.2\% & 48.3\% & 35.7\% & 32.9\% & 15.4\% & 1.1\% & 0.5\% & 7.1\% & 4.4\% & 1.6\%\\
    \oii & 26.6\% & 25.0\% & 20.0\% & 18.0\% & 5.0\% & 0\% & 1.7\% & 1.7\% & 3\% & 0\% \\
    \hline
    \end{tabular}
    \label{tab:infra}
    
\end{table*}

Table \ref{tab:parametros} summarises the parameters employed in the fitting. Those were selected for sampling all the parameter space, although restricted enough to have reasonable computation times.

\begin{table*}[]
    \centering
    \caption{Modules and parameters employed in CIGALE for the SED fittings of the emitters.}
    \begin{tabular}{c|c|l}
    \hline\hline
            &             & tau\_main: 2000.0, 6000.0, 8000.0 Myr\\
       Star formation history (SFH)  & sfh\_delayedbq & age\_main: 3000.0, 4000.0, 6000.0, 8000.0, 9000.0 Myr\\
         & & age\_bq: 10, 25, 500, 1500 Myr\\
         & & r\_sfr: 0., 0.2, 0.7, 1., 3., 5.\\
    \hline
     & & \\
    Single stelar population (SSP) & BC03 & metallicity: 0.004, 0.008, 0.02, 0.05\\
        &   \cite{byc2003}  & IMF: Chabrier\\
    \hline
    & & \\
    Nebular emission & Nebular & Ionisation parameter: -4.0, -2.0, -1.0 \\
    & & Gas metallicity: 0.004, 0.02\\
    \hline
     & & E\_BV\_lines: 0., 0.01, 0.05, 0.1, 0.3, 0.5, 0.7, 0.9, 1.1 \\
     Dust attenuation & dustat\_modified\_starbust & UV\_bump\_with: 0.0, 3.0 \\
     & \cite{calzetti2000} & Powerlaw\_slope: -1.5, -0.5, 0.0, 0.5 \\
     \hline
      & &  \\
      Dust emission & dl2014 & umin: 0.100, 1.000, 10.00, 50.00 \\
      & \cite{dl2014} & powerlaw slope: 1.0, 1.3, 1.7, 2.3, 2.7, 3.0 \\
    \hline
    \end{tabular}
    \label{tab:parametros}
\end{table*}

\begin{figure}
    \centering
    \includegraphics[width=\hsize]{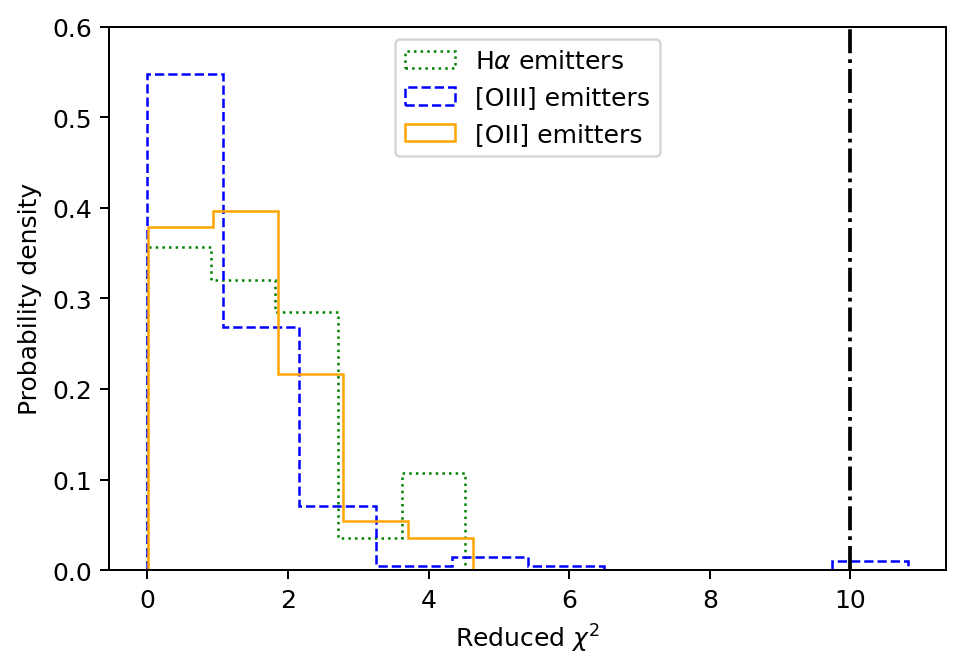}
    \caption{Histogram of the value of the reduced $\chi^2$ from the CIGALE fits. The dotted green line, the dashed~blue line, and the solid orange line represent the values of \ha, \oiii, and \oii\ emitters respectively. The vertical dash-dotted black line represents the value of $\chi^2=10$.}
    \label{fig:redchisquared}
\end{figure}

In Fig. \ref{fig:redchisquared} we show the reduced $\chi^2$ of the SED fittings obtained with CIGALE. We have marked with the dash-dotted line the value of $\chi^2=10$. This value was selected, following \cite{santos2021} and \cite{wang2020}, as the maximum tolerable for a good fit. As can be seen, only one galaxy presents a $\chi^2>10$, so it was discarded in the subsequent analysis. The rest of the galaxies of all subsamples have mean values of $\chi^2$ of 1.6, 1.5, and 1.5 for \ha, \oiii\ , and \oii\ respectively. Thus, we consider them to be reliable fits and we used them in the generation of the MS.

\begin{figure}
    \centering
    \includegraphics[width=\hsize]{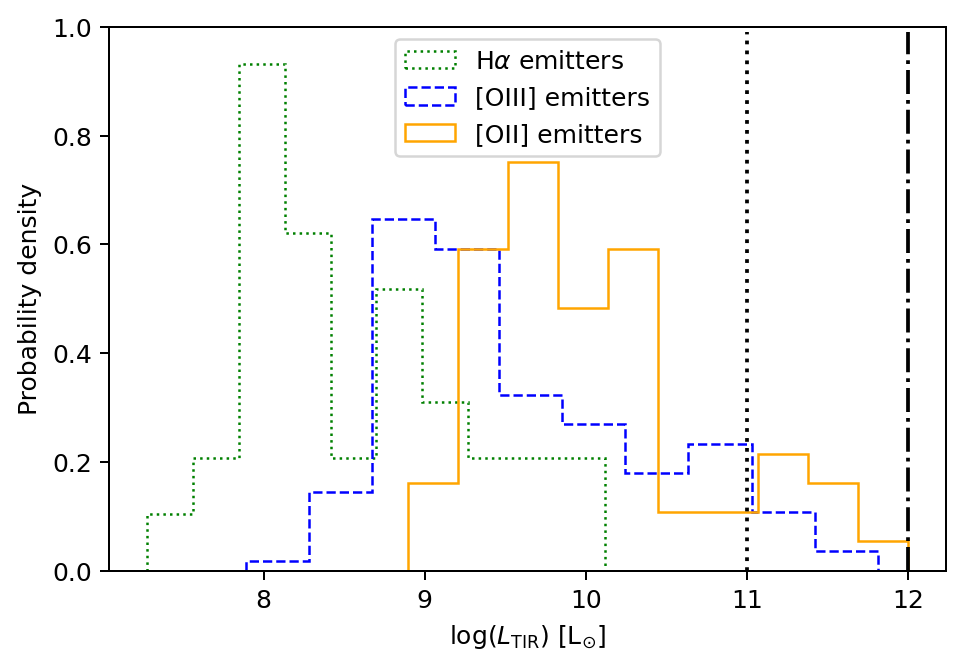}
    \caption{Histogram of the logarithm of the $L_\mathrm{TIR}$ obtained from CIGALE. Colours and line types coded as in Fig. \ref{fig:redchisquared}. The vertical dotted black line indicates the luminosity limit for LIRGs ($10^{11}$L$_{\odot}$), and the vertical dash-dotted line indicates the luminosity limit for ULIRGs ($10^{12}$L$_{\odot}$).}
    \label{fig:lumlim}
\end{figure}

From the fitting, we obtained, among others, the following parameters: bayes.stellar.m\_star\_old, which gives us the total stellar mass; bayes.dust.luminosity, which gives us the total IR luminosity; and bayes.sfh.sfr, which gives us the total SFR from the SFH fitted by CIGALE.

In Fig. \ref{fig:lumlim} we show the histogram of the total IR luminosity, extracted from CIGALE fits. The dotted line indicates the regime of luminous IR galaxies (LIRGs) at $10^{11}\mathrm{L_{\odot}}$, and the dash-dotted line indicates the regime of ultra-luminous IR galaxies (ULIRGs) at $10^{12}\mathrm{L_{\odot}}$, defined by \cite{sanders1996}. It can be seen that only a few galaxies can be classified as LIRGs, just for \oii\ (7\%) and \oiii\ (4\%), and none as ULIRGs. This indicates a low brightness in general on the IR regime for all galaxies in the subsamples.

\begin{figure}
    \centering
    \includegraphics[width=\hsize]{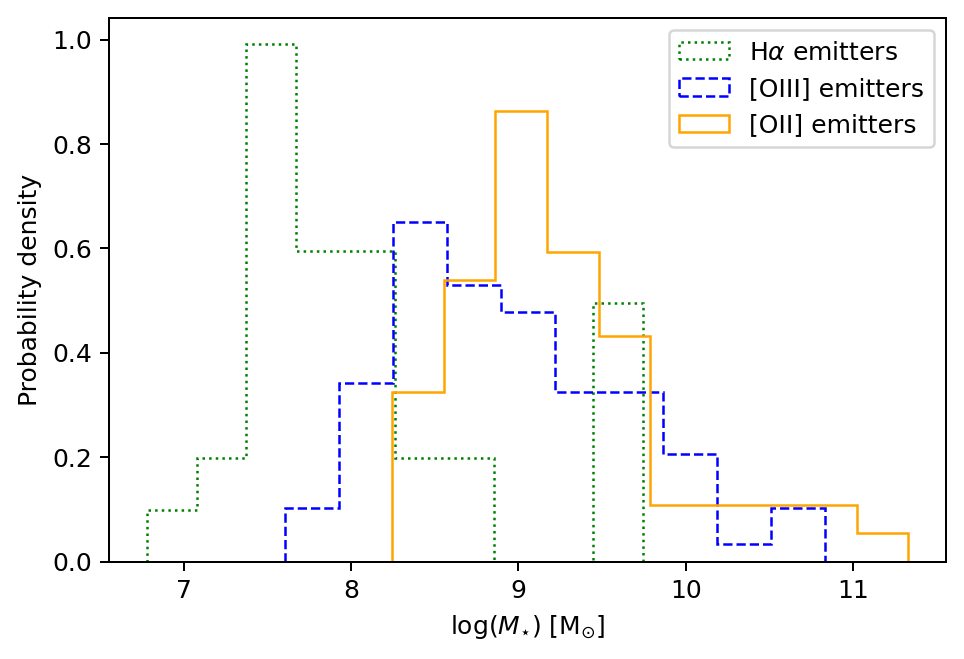}
    \caption{Histogram of the logarithm of the stellar mass obtained from CIGALE fit of the subsets emitters. Colours and line types are coded as in Fig. \ref{fig:redchisquared}.} 
    \label{fig:masa}
\end{figure}

In Fig. \ref{fig:masa} we use a histogram to represent the stellar mass of our emitters. The median values for the logarithm of the masses (in $M_{\odot}$) are 7.67, 8.85, and 9.11 for the emitters \ha, \oiii, and \oii\ , respectively. Most of the emitters (>90\%) are below $10^{10}$M$_{\odot}$, with galaxies reaching as low as $10^{7}$M$_{\odot}$. Taking this into account, we can consider most of our emitters to be found in the low-mass regime. This is a direct consequence of the OTELO Survey being a pencil-beam survey, which generates a bias towards low-mass galaxies (see, for example, \citealt{impey1997}, \citealt{moster2011} and \citealt{cedres2021}). Taking into account also that the median significative mass for LIRGs is about $6.3\times10^{10}$M$_{\odot}$ (\citealt{paspaliaris2021}), it is reasonable to expect few LIRGs in our sample, and no ULIRGs at all, as is seen in Fig. \ref{fig:lumlim}.

\section{The main sequence of galaxies} \label{sec4}

\subsection{Star formation rate}

The SFR can be estimated from hydrogen recombination lines of
H {\sc ii} regions (see \citealt{kennicutt1998} and references therein), since they scale linearly with ionizing photons produced by recently formed OB stars. This SFR estimate is actually an average of the star formation history over the last 10\,Myrs \cite[see][for a formal description]{Cervinoetal16}. In the case of the optical range, when Balmer lines are not available, some intense collisional forbidden lines have,
as in \oii\ \cite[e.g.][]{kewley2004} or \oiii\ \cite[e.g.][]{suzuki2016}, been proposed as a SFR estimate, although, since such lines do not scale linearly with the ionizing flux of OB stars (\citealt{Villaverde10}), extra assumptions would be needed about the ionization structure and flux of the galaxy, and to include a metallicity dependence in their use as a SFR tracer.

Another alternative is to use the total infrared luminosity ($L_\mathrm{TIR}$) as an estimate of the SFR, although in this case it produces a value averaged over the last 100 Myrs, a time span much larger than the one obtained from emission lines. Using
$L_\mathrm{TIR}$ will give us the possibility of employing all the subsets, sampling the whole range of redshifts, and therefore different windows on the age of the Universe.

The SFR from the $L_\mathrm{TIR}$ can be calculated following \cite{kennicutt2012}:
\begin{equation}
    \log(SFR)=\log(L_{TIR})-43.41,
    \label{eq_pop}
\end{equation}
where the SFR is expressed as units of solar masses per year and $L_\mathrm{TIR}$ has units of ergs per second.

\begin{figure}
    \centering
    \includegraphics[width=\hsize]{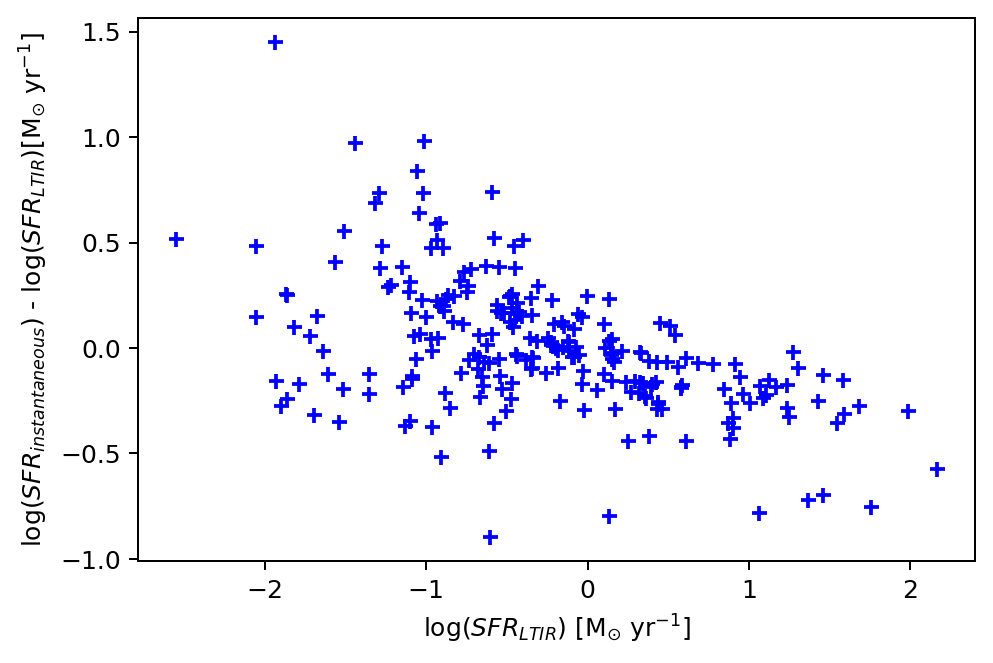}
    \caption{Difference between the logarithm of $SFR_\mathrm{instantaneous}$ and the logarithm of $SFR_\mathrm{LTIR}$ as a function of the logarithm of $SFR_\mathrm{LTIR}$ for all the emitters in the sample.}
    \label{fig:oii_delta}
\end{figure}

The values for $L_\mathrm{TIR}$ are part of the output of CIGALE.
We can compare this SFR from $L_\mathrm{TIR}$ (thereafter $SFR_{LTIR}$) with the SFR given by the SFH of the CIGALE fit.
In Fig.\ref{fig:oii_delta}, we represent the difference between the logarithm of $SFR_\mathrm{LTIR}$ and the logarithm of the instantaneous SFR from the derived SFH, $SFR_\mathrm{instantaneous}$, as a function of the logarithm of $SFR_\mathrm{LTIR}$.

\begin{figure}
    \centering
    \includegraphics[width=\hsize]{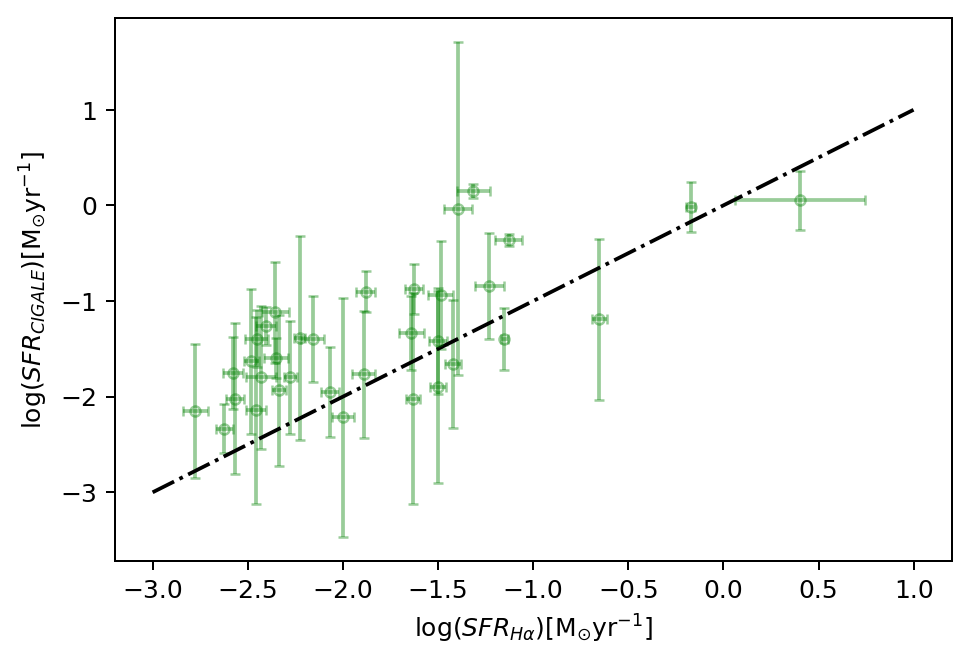}
    \caption{Logarithm of the instantaneous SFR obtained by CIGALE vs the logarithm of the SFR from \ha\ emission line. The dash-dotted line represents 1:1 equivalence.}
    \label{havsha}
\end{figure}

We should expect a constant value of the difference, with some dispersion and an offset around 0. However, there is a clear dependence on the $SFR_{LTIR}$. Such a dependence indicates that galaxies with very low emission in the IR part of the spectrum present lower $SFR_{LTIR}$ values than expected. This means that the determination of $L_\mathrm{TIR}$ is not accurate, at least for low-luminosity galaxies. This could be due to a problem with obtaining the dust luminosity from CIGALE without enough information in the IR part of the SED. In fact, most of the emitters are largely devoid of IR data, with only a select number of galaxies having data at MIPS24\,$\mu$m and at longer wavelengths, as can be seen in Table \ref{tab:infra}. It is possible that the lack of IR constraints generates problems when calculating $L_\mathrm{TIR}$ for CIGALE. This behaviour has been reported for other models of SED fittings (\citealt{hunt2019}) but not on CIGALE. Taking this into account, we used the SFR given directly by CIGALE from the SFH. As previously seen, low-mass galaxies have low dust content, so $L_\mathrm{TIR}$ is also low. For this reason, those galaxies are not detected at IR wavelengths, and $L_\mathrm{TIR}$ in general is not a good indicator of the SFR. On the other hand, the CIGALE SFR we are going to use is bayes.sfh.sfr, which corresponds to the instantaneous SFR (\citealt{cigale}), while the SFR from $L_\mathrm{TIR}$ provides a measure of the SFR obscured by the extinction (\citealt{calzetti2000}, \citealt{kennicutt2009}). This instantaneous SFR should be pretty similar to the one obtained by the emission of the emission lines. In Fig. \ref{havsha} we have represented the SFR derived from CIGALE against the SFR obtained by \ha\ emission line (from \citealt{marina2019}) for our subsample of \ha\ emitters. CIGALE derived SFRs have large uncertainties, which generates a large scatter, and there are few galaxies with \ha\ derived SFRs. However, taking into account the uncertainties, there is a correspondence between the two quantities.

\subsection{Fitting of the main sequence}

\begin{figure}
    \centering
    \includegraphics[width=\hsize]{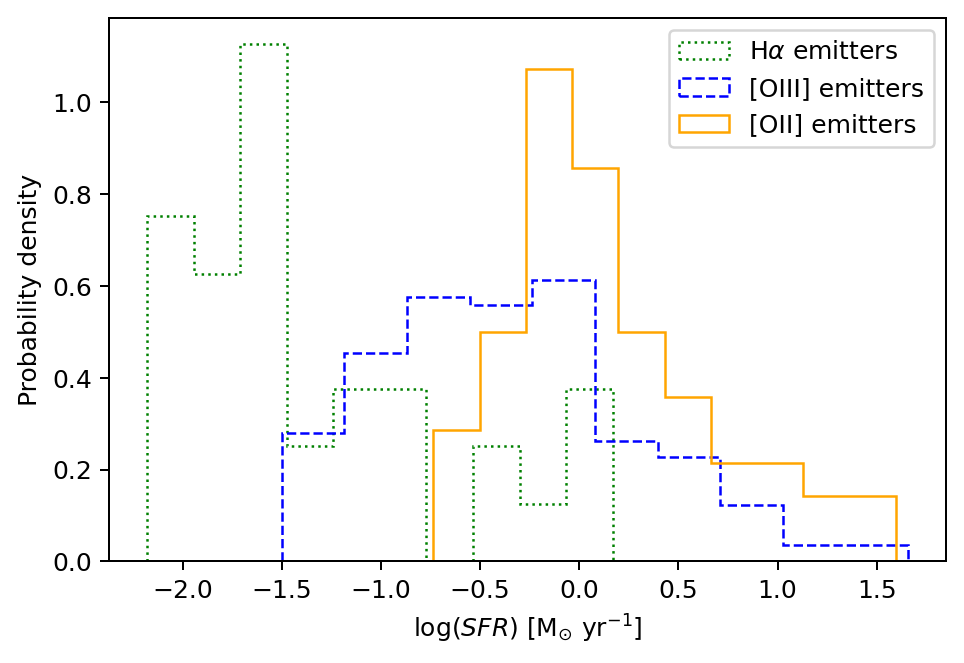}
    \caption{Histogram of the logarithm of the SFR from CIGALE for all the selected emitters. Colours and line types are coded as in Fig. \ref{fig:redchisquared}.}
    \label{fig:sfr-cigale}
\end{figure}
\begin{figure}
    \centering
    \includegraphics[width=\hsize]{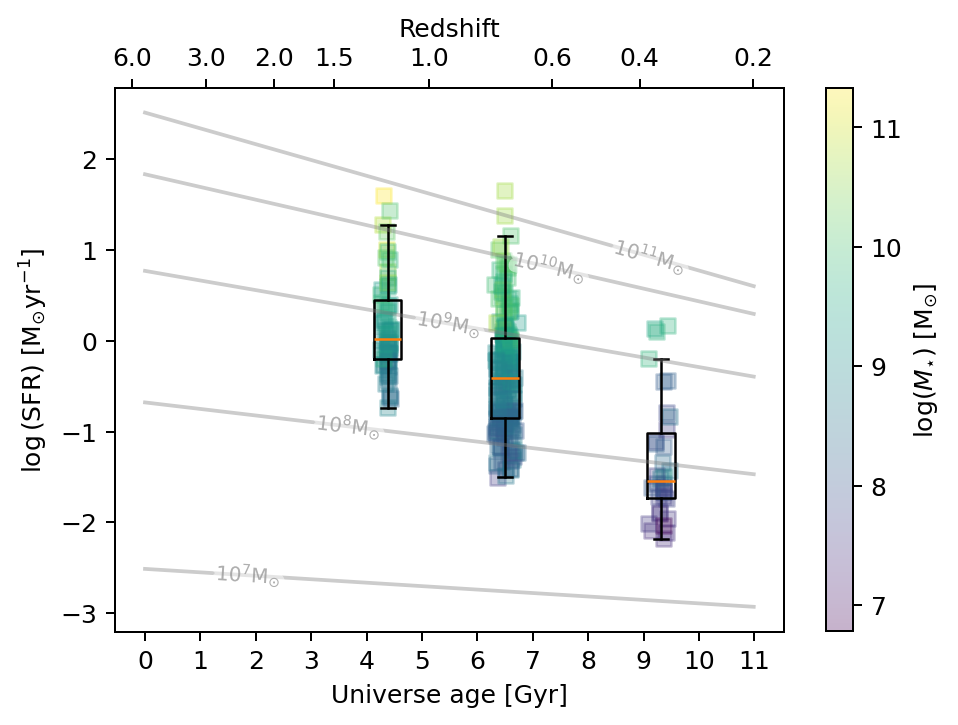}
    \caption{Logarithm of the SFR as a function of the age of the Universe. The median value is indicated by the orange line. The boxes are defined by the first quartile as the bottom line, and the third quartile as the upper line. The whiskers are calculated as 1.5 times the inter-quartile range. The masses of the emitters are color-coded following the color bar on the right. The grey lines are from eq. \ref{eq_pop} from \cite{popesso2023}.}
    \label{fig:sfrbigotes}
\end{figure}

 Taking into account that the parameters presented in eq. \ref{eclee} may depend on the age of the Universe, we can rewrite this equation, following \cite{popesso2023}, as
\begin{equation}
    \log SFR(M_*,t)=a_0+a_1t-\log\left(1+(M_*/10^{a_2+a_3t})^{-a_4}\right),
    \label{pop2}
\end{equation}
where $a_0=2.68$, $a_1=-0.186$, $a_2=10.83$, $a_3=-0.0729$, $a_4$ as $\gamma$, where $a_1$ and $a_3$ go with inverse time units, and $t$ as the age of the Universe, expressed in gigayears. Employing the parameters $a_2$ and $a_3$, it is possible to parameterize the value of the turnover mass over time.

In Fig. \ref{fig:sfr-cigale} we have represented the histogram of the logarithm of the SFR for all the emitters in the subsamples. It can be seen that the distribution of the emitters is different, with \ha\ emitters presenting lower values of the SFR. A similar effect is seen in Fig. \ref{fig:sfrbigotes}, where we show the logarithm of the SFR as a function of age of the Universe. The grey lines are obtained from eq. \ref{pop2} and represent the evolution of the SFR of galaxies with fixed stellar masses across time. Although there is a large dispersion, it can be seen that there is an evolution of the SFR with time, with median values of the SFR larger for a lower age of the Universe. This implies a different distribution of the SFR for the \ha\ emitters, as can be seen in Fig. \ref{fig:sfr-cigale}, with lower SFR values compared with the other two subsamples. However, it should be noted that we are not detecting low-SFR galaxies at a lower age of the Universe due to its low brightness (the so-called Malmquist effect).

\begin{figure}
    \centering
    \includegraphics[width=\hsize]{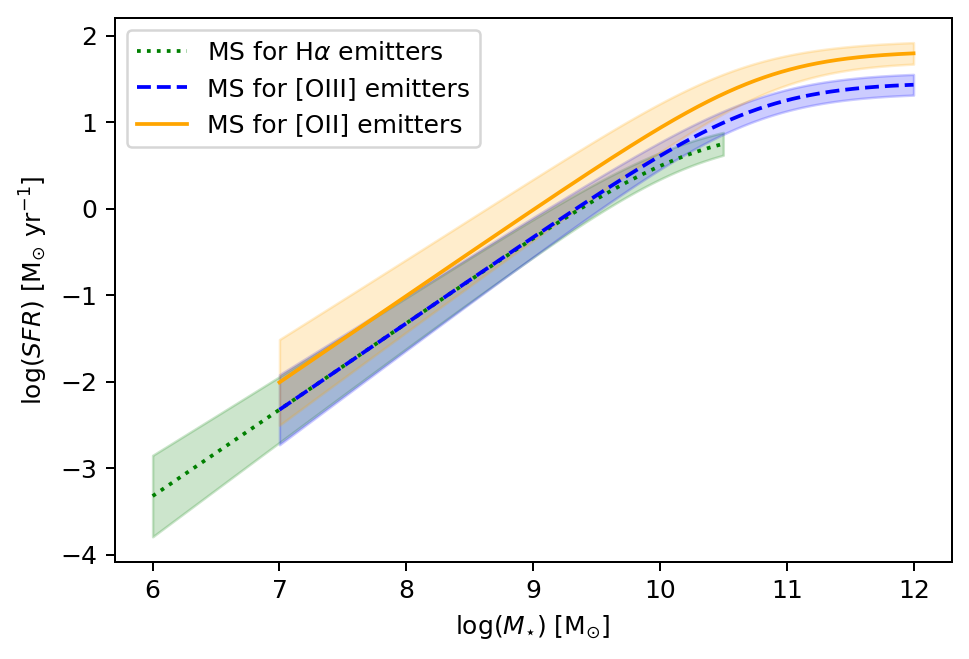}
    \caption{Fitted MS for the different emitters in the sample following eq. \ref{eclee}. The dotted green line, dashed blue line, and solid orange line represent the fit for \ha, \oiii, and \oii\ emitters, respectively. The shaded areas represents the propagation of $1\sigma$ uncertainties.}
    \label{ms_total}
\end{figure}

In Fig. \ref{ms_total} we represent the logarithm of the SFR as a function of the stellar mass for the \ha, \oiii, and \oii\ emitters.  The data were fitted to eq. \ref{eclee} following a Monte Carlo procedure. A random value was assigned to each data point within their error bars, and then the entire data were fitted using a least squares minimization algorithm based on the Levenberg–Marquardt method. This process was repeated $10^5$ times. For each parameter fitted, it was assumed that the median of all the fittings was their main value, and as their uncertainty the standard deviation was selected. Taking into account the lack of galaxies at higher masses ($\log(M_*/\ensuremath{M_{\odot}})>10$, it was not possible to determine the value of $SFR_\mathrm{max}$ without increasing the uncertainties for the rest of the parameters. Then, we used the value given by \cite{popesso2023}. The results are summarised in table \ref{ajuste}, where we have also included the value of the age of the Universe used in each subsample.

\begin{table*}[]
    \centering
    \caption{Parameters and fitting results of eq. \ref{eclee} for the MS.}
    \begin{tabular}{cccccc}
    \hline
    \hline
        Source & Age of the Universe & $\log (SFR_{max})$ & $\log(M_0)$ & $\gamma$ & $\log(M_0)$ binned\\
        & [Gyr] & [M$_{\odot} yr^{-1}$] & [M$_{\odot}$] & & [M$_{\odot}$]\\
        \hline
         \ha & 9.31 & 0.95 (fixed) & 10.27$\pm$0.14 & 0.96$\pm$ 0.10 & 10.14$\pm$0.13\\
         \oiii & 6.50 & 1.46 (fixed)& 10.78$\pm$0.10 & 1.20$\pm$0.10& 10.52$\pm$0.11 \\ 
         \oii & 4.38 & 1.82 (fixed)& 10.83$\pm$0.26 & 1.05$\pm$0.19 & 10.85$\pm$0.13 \\
         \hline
    \end{tabular}
    \label{ajuste}
\end{table*}

\begin{figure}
    \centering
    \includegraphics[width=\hsize]{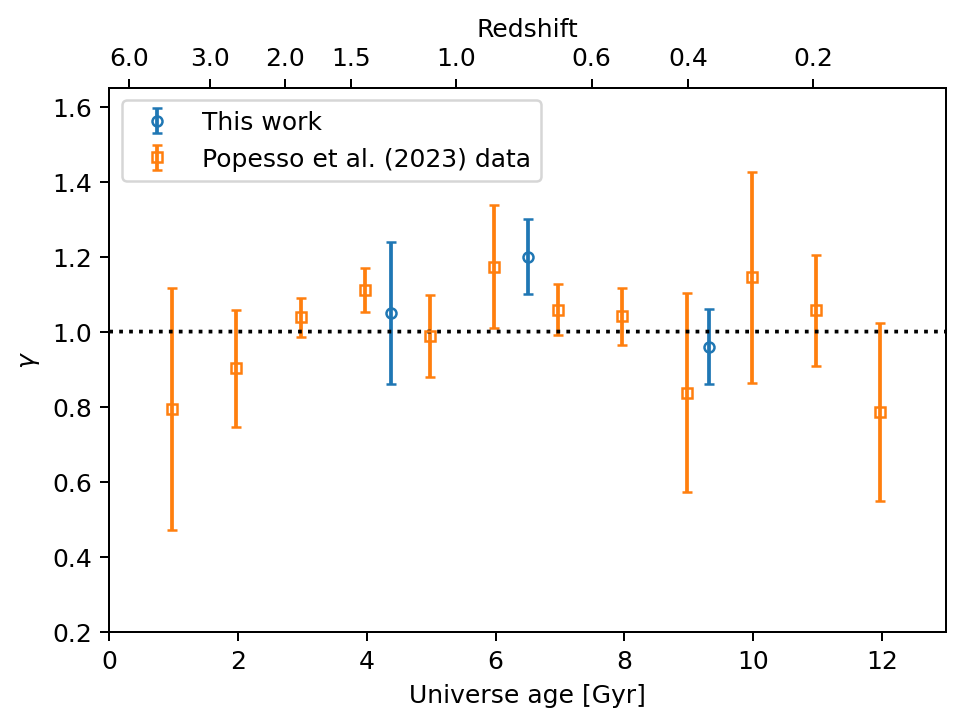}
    \caption{Slope of the low-mass regime, $\gamma$, as a function of the age of the Universe. Symbols are indicated in the legend.}
    \label{gamma}
\end{figure}

In Fig. \ref{gamma} we represent the variation in $\gamma$ as a function of the age of the Universe. \cite{lee2015} suggested a possible evolution of $\gamma$ with the age of the Universe, with higher values at higher redshifts. Our results may indicate a different trend if taken alone. However, if we include the data from \cite{popesso2023} in the context, the general trend is compatible with a constant value of $\gamma=1$, as stated in \cite{popesso2023}.

\begin{figure}
    \centering
    \includegraphics[width=\hsize]{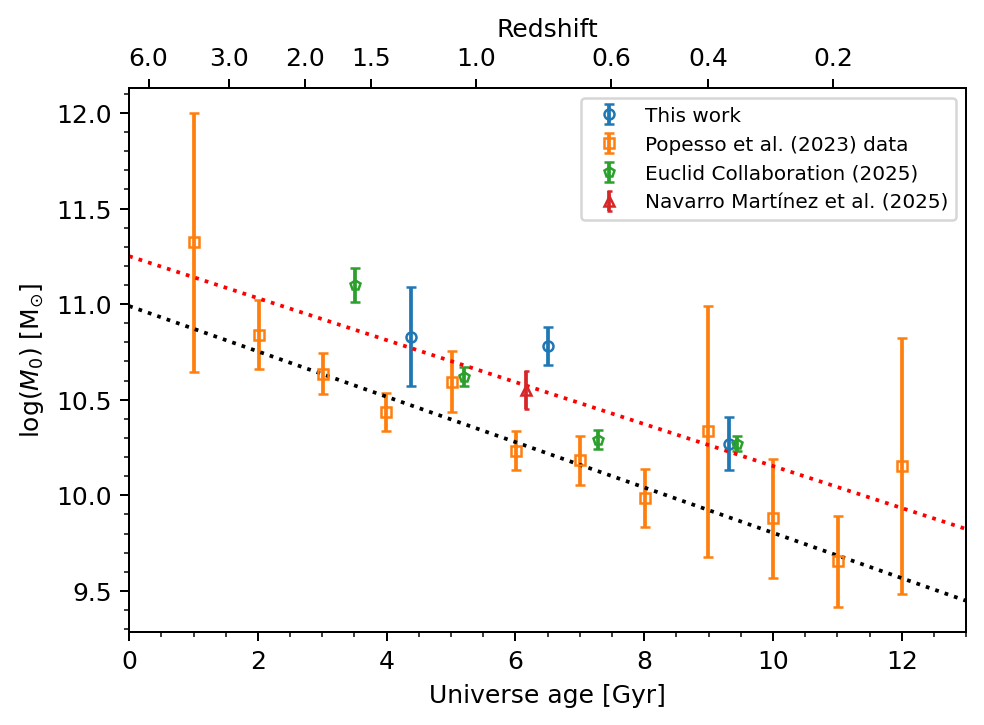}
    \caption{Turnover mass as a function of the age of the Universe. Symbols are indicated in the legend. The dotted black line represents the fit given by \cite{popesso2023}. The dotted red line represents the fit obtained employing the data from \cite{euclid2025}, \cite{rocio2}, and this work.}
    \label{m0}
\end{figure}

In Fig. \ref{m0} we represent the turnover mass as a function of the age of the Universe. We have included data from \cite{popesso2023}, \cite{euclid2025}, and \cite{rocio2}. Our results are within the uncertainties presented in the previous works and follow the general trend of lower values of $\log(M_0)$ with the age of the Universe, as stated in \cite{lee2015}. However, our data consistently present a higher value of turnover mass for all samples compared to the data from \cite{popesso2023}.

\begin{table*}[]
    \centering
    \caption{Results of the linear fitting of $\log(M_0)$ with the age of the Universe.}
    \begin{tabular}{cc|c}
    \hline
    \hline
       Source & $a_2$ & $a_3$\\
    \hline
       \cite{popesso2023} fit  & $10.99\pm0.08$ & $-0.118\pm0.013$  \\
        This work only fit & $11.68\pm0.37$ & $-0.146\pm0.051$ \\
        \cite{euclid2025} +& & \\  \cite{rocio2} +  & $11.25\pm0.19$ &  $-0.11\pm0.02$\\this work fit & &  \\
    \hline
    \end{tabular}
    \label{constantes}
\end{table*}

In Table \ref{constantes} we summarise the fits to a straight line of the turnover mass with the age of the Universe in the form $\log(M_0)=a_2+a_3t$, where $a_2$ and $a_3$ are the same as in eq. \ref{pop2}. We can observe that the main difference lies in the intercept value, $a_2$, which is larger for our data, and slightly larger for the combination of our data with \cite{euclid2025} and \cite{rocio2} data. However, the differences are within the range of the uncertainties.

Nevertheless, we can explore why the results from \cite{popesso2023} present a somewhat lower value for the turnover mass compared to our data. If we consider this difference to be a physical one, that means that the sSFR of low-mass galaxies is lower for our emitters when compared with the galaxies from \cite{popesso2023}. This means that our low-mass emitters, the bulk of our sample, are less efficient in forming stars than the galaxies of \cite{popesso2023}. However, from \cite{beliletter}, it is clear, at least for \ha\ and \oii\ emitters, that they present a sSFR compatible with so-called normal SFGs, so there has to be another reason for the difference. 
It is important to note that the data from \cite{popesso2023} and our data are very different. Our sample reaches very low SFR values ($\log(SFR)<-1.0$\,[M$_\odot$yr$^{-1}$] for \ha\ and \oiii\ subsamples), while the full dataset by \cite{popesso2023} stops at $\log(SFR)\sim-1.0$\,[M$_\odot$yr$^{-1}$]. The \cite{popesso2023} data are a compilation of different obtained SFRs and different selection processes. For example, our data are selected only by emission line detection, while in the \cite{popesso2023} compilation there is only one instance of selection by \ha\ emission line between 28 different subsets. Moreover, the SFR indicators are also different, ranging from SED fits (like ours) to a combination of NUV+IR, radio, emission lines, or IR, among others. Although \cite{popesso2023} performed extensive work to homogenize the different datasets, there is a clear difference in the methods employed in the work presented there, and some dispersion may have been introduced in the final determination of the evolution of MS over time (see, for example, Fig. A2 from \citealt{popesso2023}, where a comparison between different determinations of the SFR is presented). On the other hand, our results are pretty close to those of \cite{euclid2025}, which are based on a homogeneous dataset, although it is a survey that covers a larger surface of the sky, compared to the OTELO survey ($\sim$63\,deg$^2$ for \citealt{euclid2025}, and only 56\,arcmin$^2$ for OTELO). It can be assumed then that part of the difference may come from the homogeneity versus inhomogeneity of the datasets.

Furthermore, the differences in the methods of obtaining the MS cannot be overlooked either, because they may play a role in the observed difference, as \cite{popesso2023} pointed out. In our case, we did not employ the binning in mass while generating the MS, unlike previous works (i.e. \citealt{popesso2023}, \citealt{daddi2022}, \citealt{lee2015}, or \citealt{euclid2025}). 

\subsection{Binning the data for the MS}

\begin{figure}
    \centering
    \includegraphics[width=\hsize]{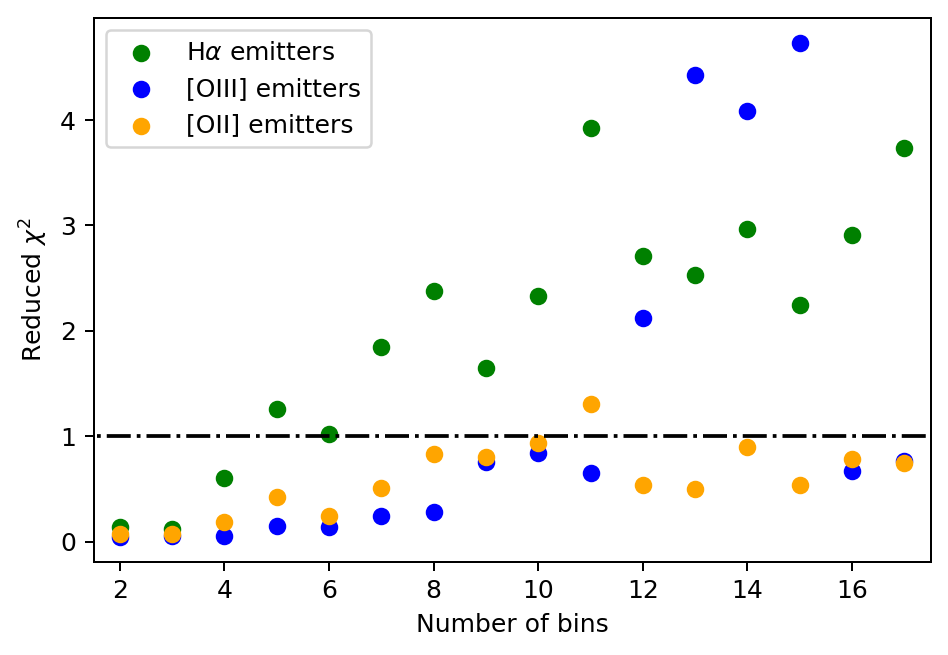}
    \caption{Reduced $\chi^2$ of the fit as a function of the number of bins for each emitter. Symbols are indicated in the legend. The horizontal dash-dotted line represents a value of $\chi^2=1$.}
    \label{fig:chi_fit}
\end{figure}

\begin{figure}
    \centering
    \includegraphics[width=\hsize]{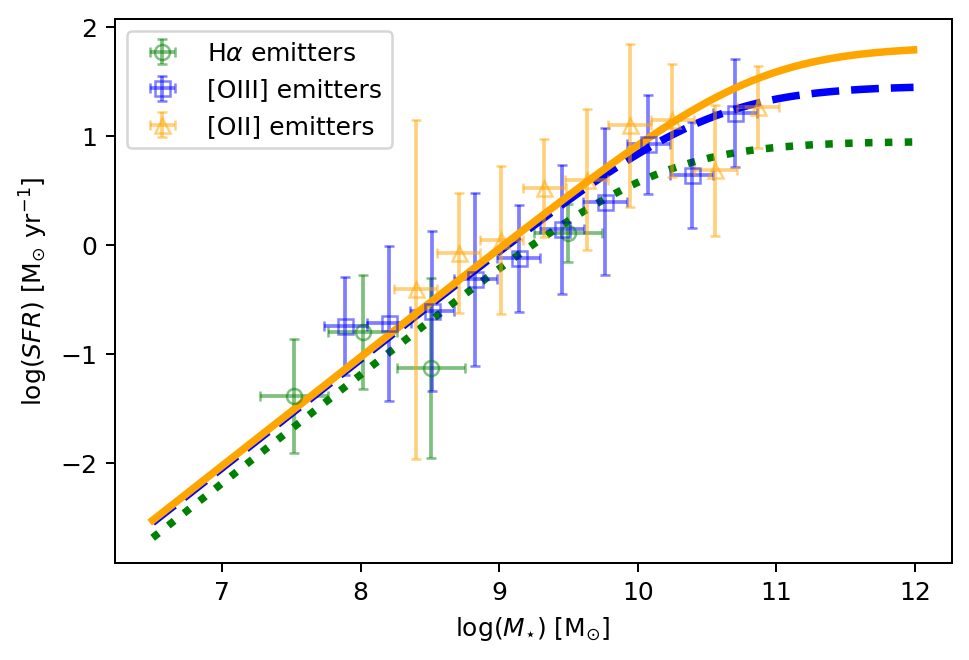}
    \caption{Binned MS for the emitters. Symbols are indicated in the legend. The dotted green line, the dashed blue line, and the solid orange line represent the fitted MS for the \ha, \oiii, and \oii\ emitters, respectively.}
    \label{fig:allbin}
\end{figure}

In order to make a comparison with the rest of the data, we chose to bin our data. Taking into account that the sample variable in the MS is the stellar mass, we decided to bin into equal-size bins in $\log(M_*)$. To select the size (and therefore the number) of bins that best sample our dataset, we generated fits of the MS for 2 bins to 17 bins.% where the mean number of galaxies per bin is reduced to 1. 
We then calculated the weighted average of the SFR in each bin. The weight given to each emitter was the inverse of the quadrature sum of the uncertainties in both the mass and the SFR. In this case, we fixed both values of $\gamma=1$ and $\log(SFR_{max})$. As a metric, to evaluate the goodness of each fit, we employed the reduced $\chi^2$, and selected as the best fit the one with the closest value to 1. These fits gave us 6, 10, and 10 bins, with a mean value of 6, 18, and 6 galaxies per bin, for \ha, \oiii, and \oii\ , respectively. In \ref{fig:chi_fit} is shown the variation in $\chi^2$ with the different number of bins employed. 
In Table \ref{ajuste} are indicated the values obtained in the turnover mass for each subset of emitters and in Fig. \ref{fig:allbin} we represent the MS obtained by binning the data.  

\begin{figure}
    \centering
    \includegraphics[width=\hsize]{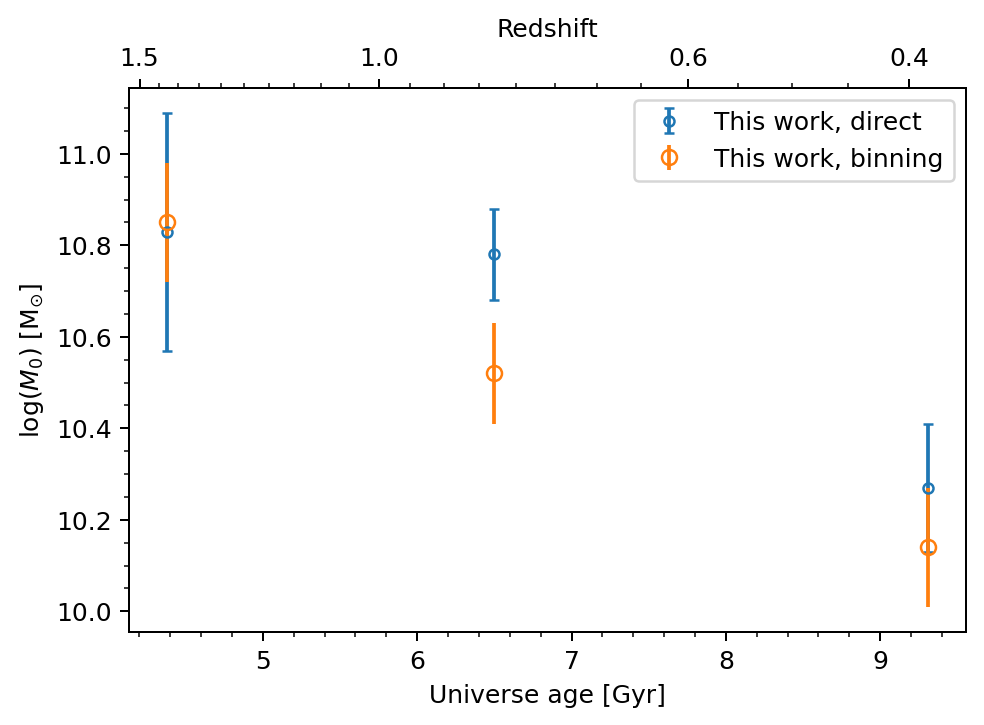}
    \caption{Turnover mass for our data as a function of the age of the Universe. The open blue circles were obtained by the direct fitting of all the emitters. The open orange circles were obtained by employing the binned data.}
    \label{fig:bineo}
\end{figure}

In Fig. \ref{fig:bineo} we show the comparison of the turnover mass obtained from the direct fit and the turnover mass obtained by binning the data, represented by open blue circles and open orange circles, respectively. The results of the two methods seem to be compatible, although for \ha\ and \oiii\ the values of the binned turnover mass are lower. The binning may explain some of the differences between our results and those of \cite{popesso2023}, and the method used to generate the MS may play a role in the determination of the turnover mass. However, the variations seem to be smaller than the ones presented in Fig. \ref{m0}, and our non-binned results match better with those of \cite{euclid2025} and \cite{rocio2}, which were obtained by binning the data. It should be taken into account that the compilation from \cite{popesso2023} has a heterogeneous sample of data from different authors and methods. Even with careful conversions made to the data to homogenise them, there is variability that results in values of $\log(M_o)$ with a high uncertainty and this may have an influence on the difference obtained in this work.

\section{Conclusions} \label{sec5}
In this work, we have presented a study in the evolution of the MS of ELGs from redshift 0.38 to 1.43, using three subsamples of ELGs (\ha, \oiii, and \oii\ lines) from OTELO Survey. We confirmed the results from \cite{otelo}, where the use of only colour-magnitude relationships techniques in the selection of ELGs will exclude a larger ($\sim$29\%) number of emitters.

Employing the CIGALE code, we fitted a SED for each galaxy of the subsamples. We selected a value of $\chi^2<10$ as a limit to consider a fit as a ‘good’ fit, with mean values for our sample of 1.6, 1.5, and 1.5 for \ha, \oiii, and \oii\ , respectively.
From the SED fitting, we obtained the stellar mass, $L_\mathrm{TIR}$, and SFR of each emitter.

As expected from the previous works (\citealt{marina2019}, \citealt{bongio2020}, \citealt{cedres2021}\ , and \citealt{rocio2021}), almost all of the emitters from the three subsets are low-mass galaxies ($<10^{10}\mathrm{M_{\odot}}$). This, as a consequence, is translated as a small number of LIRGs, and a total absence of ULIRGs in our samples.

We found that the determination of $SFR_{LTIR}$ was not reliable due to a lack of IR constraints in the fitting of the SEDs. The low stellar masses of our emitters also implied low dust masses, and consequently low IR brightness. This means that $L_\mathrm{TIR}$ is not a good indicator of the SFR for low-mass galaxies.  For this reason, we decided to use the SFR directly from the SFH fitted by CIGALE.% which is an average of present and past star formation, and, therefore a better estimation of the total SFR of the galaxies.

We generated the MSs for the \ha, \oiii, and \oii\ emitters, and fitted them to the functional form proposed by \cite{lee2015}. From these fits, we derived the slope value of the low-mass regime, $\gamma$, and the turnover mass, $\log(M_0)$. Due to our low number of high-mass galaxies ($>10^{10}\mathrm{M_{\odot}}$), we decided to use the value for $SFR_\mathrm{max}$ given by \cite{popesso2023}. The results for $\gamma$ were compatible with a constant value of 1 throughout the age of the Universe, as suggested in \cite{popesso2023}. We also find an evolution of the $\log(M_0)$ with the age of the Universe. However, in this case, the values of $M_0$ obtained in our fits are larger than the values given in \cite{popesso2023} for all the redshifts sampled. This trend is also present when including data from \cite{euclid2025} and \cite{rocio2}. However, the general slope of the variation in $\log(M_0)$ is compatible between our results and the fit from \cite{popesso2023}. 
When instead of using all emitters, we generated a binned average of the SFR, we obtained values similar to our not-binned results (although still somewhat lower for \ha\ and \oiii\ ). However, it should be noted that the data from \cite{euclid2025} and \cite{rocio2} are closer to our results than to those of \cite{popesso2023}. This may be due to the use of more homogeneous data in \cite{euclid2025}, \cite{rocio2}, and this work, compared to the compilation from different sources presented in \cite{popesso2023}. We can assume, then, that at least some of the differences are attributable to the uncertainties introduced by the different methodologies employed.

\begin{acknowledgements}
The authors wish to thank the anonymous referee for her/his feed-back and useful suggestions, which have contributed to significantly improving the manuscript.
B.C. and J.C. ~acknowledge the support of the Spanish Ministry of Science, Innovation and Universities through the project PID--2021--122544NB--C41.
A.B., M.I.R., and M.S.P. acknowledge the support of the Spanish Ministry of Science, Innovation and Universities through the project PID–2021–122544NB–C43.
M.C. acknowledges support from the Spanish grant PID2022-136598NB-C33 funded by MCIN/AEI/10.13039/501100011033 and by “ERDF A way of making Europe”.
J.A.D. acknowledges support from the PAPIIT-DGAPA\_UNAM grant IN116325.
J.N. acknowledge the support of the National Science Centre, Poland, through the SONATA BIS grant 2018/30/E/ST9/00208 and the support of the Polish National Agency for Academic Exchange (NAWA) Bekker grant BPN/BEK/2023/1/00271, and the kind hospitality of the IAC.
M.G.O. acknowledges financial support from the State Agency for Research of the Spanish MCIU through Center of Excellence Severo Ochoa award to the Instituto de Astrof\'isica de Andaluc\'ia CEX2021-001131-S funded by MCIN/AEI/10.13039/501100011033, and from the grant PID2022-136598NB-C32 “Estallidos8”. MGO also acknowledges the support by the project ref. AST22\_00001\_Subp\_11 funded from the EU – NextGenerationEU,  PPCC Junta de Andaluc\'ia.

J.G. acknowledges the support of the Spanish Ministry of Science, Innovation and Universities through the projects PID--2021--123417OB--I00 and PID--2024--157374OB-I00.

M.P. acknowledges the support from the Spanish Ministry of Science, Innovation and Universities through projects PID--2022--140871NB--C21 and PID--2024--162972NB--I00, and the State Agency for Research of the Spanish MCIU through the Center of Excellence Severo Ochoa award to the Instituto de Astrof\'isica de Andaluc\'ia (CEX2021-001131-S funded by MCIN/AEI/10.13039/501100011033).

M.A.L.L. acknowledges support from the Spanish grant PID--2021-–123417OB--I00, and the Ram\'on y Cajal program funded by the Spanish Government (RYC2020--029354--I).

J.I.G-S. acknowledges the support of the Spanish Ministry of Science, Innovation and Universities through the project PID--2021--122544NB--C44.

      B.C. wishes to thank Carlota Leal \'Alvarez by her support during the development of this paper.  

This article is based on observations made with the Gran Telescopio Canarias (GTC) at Roque de los Muchachos Observatory of the Instituto de Astrof\'isica de Canarias on the island of La Palma. 

This study makes use of data from AEGIS, a multi-wavelength sky survey conducted with the Chandra, GALEX, Hubble, Keck, CFHT, MMT, Subaru, Palomar, Spitzer, VLA, and other telescopes, and is supported in part by the NSF, NASA, and the STFC. 

Based  on  observations  obtained  with  MegaPrime/MegaCam,  a  joint  project  of the  CFHT  and CEA/IRFU, at the Canada--France--Hawaii Telescope (CFHT) which is operated by the National Research Council (NRC) of Canada, the Institut National des Science de l'Univers of the Centre National de la Recherche Scientifique (CNRS) of France, and the University of Hawaii.  This work is based in part on data products produced at Terapix available at the Canadian Astronomy Data Centre as part of the Canada-France-Hawaii Telescope Legacy Survey, a collaborative project of NRC and CNRS. 

Based on observations obtained with WIRCam, a joint project of CFHT, Taiwan, Korea, Canada, France, at the Canada--France--Hawaii Telescope (CFHT), which is operated by the National Research Council (NRC) of Canada, the Institute National des Sciences de l'Univers of the Centre National de la Recherche Scientifique of France, and the University of Hawaii. This work is based in part on data products produced at TERAPIX, the WIRDS (WIRcam Deep Survey) consortium, and the Canadian Astronomy Data Centre. This research was supported by a grant from the Agence Nationale de la Recherche ANR-07-BLAN-0228. 
\end{acknowledgements}


\bibliographystyle{aa} 
\bibliography{biblio} 

\end{document}